\documentclass{eas}
\usepackage{graphicx}
\usepackage{subfigure}
\begin{document}

\title{Determining properties of the Antennae system - Merging ability for restricted N-body} 
{\author{Hanns P. Petsch$^1$}
\author{Christian Theis}}
\runningtitle{Merging ability for restricted N-body}
\address{Institut f\"ur Astronomie, Universit\"at Wien, T\"urkenschanzstra\ss{}e 17, 1180 Wien, Austria}
%
%
\begin{abstract}
Motivated by the closest major merger, the Antennae Galaxies (NGC\,4038/4039), we want to improve our genetic algorithm based modeling code {\sc Minga} (Theis \cite{theis99}). The aim is to reveal the major interaction and galaxy parameters, e.g.\,orbital information and halo properties of such an equal mass merger system. Together with the sophisticated search strategy of {\sc Minga}, one needs fast and reliable models in order to investigate the high dimensional parameter space of this problem.
Therefore we use a restricted {\sl N}-body code which is based on the approach by Toomre \& Toomre (\cite{toomre72}), however with some refinements like consistent orbits of extended dark matter halos. Recently also dynamical friction was included to this code (Petsch \cite{petsch07}). While a good description for dynamical friction was found for mass ratios up to $q=1/3$ (Petsch \& Theis \cite{petsch08}), major merger systems were only imperfectly remodeled. Here we show recent improvements for a major merger system by including mass-loss and using NFW halos.\\
\end{abstract}
\maketitle
\section{Introduction}
The Antennae System (NGC\,4038/4039) is the closest observable major merger. It serves as a prototype for interacting galaxies and tidally induced activity. Current high precision observations over a large range of wavelengths give insight to the dynamical and morphological properties. 
If one wants to perform state-of-the-art high resolution self-consistent models, a reliable set of orbital and structural parameters is required.
Due to the high dimensional parameter space of this problem, one needs fast and reliable models as well as a sophisticated strategy to pick out the best one. Therefore we use the program {\sc Minga} (Theis \cite{theis99}), where an improved restricted {\sl N}-body code is coupled to a genetic algorithm. It was shown in Theis \& Kohle (\cite{theis01}), that it is possible to reproduce the large scale features of an observed system with this method. A similar approach was successfully used by Wahde \& Donner (\cite{wahde01}) and Ruzicka \etal (\cite{ruzicka07}).\\


\section{Method}
The introduction of dynamical friction in our restricted {\sl N}-body code allowed us to model close interactions and late stages of merging. Up to a mass ratio of $q=1/3$, the orbital decay and the basic tidal features of a satellite merging with an isothermal halo can be reliably reproduced (Petsch \& Theis \cite{petsch08}). While a mass and distance dependent description for the Coulomb logarithm together with an orientation correction of the frictional force was applicable for these minor mergers, it did not perfectly work for a major merging system.

In this previous work we have focused on the determination and parameterisation of the Coulomb logarithm $\ln \Lambda$. We found the best description of the dynamical friction using a modified Chandrasekhar approach (Eq.\,\ref{petsch_eq_4}).

\begin{equation}
 \frac{d \vec{v}_M}{dt} = - F(v_M,\sigma) C_f \frac{\rho M}{v_M^2} \left(\hat{v}_M \cos{(\beta)}  + \hat{e}_\bot \sin{(\beta)} \right) \ln \Lambda 
\label{petsch_eq_4}
\end{equation}
\begin{equation}
 \ln \Lambda=\ln \left[1 + \frac{M_{\mathrm{halo}}(r_M)}{M} \right]
\label{petsch_eq_6}
\end{equation}
\noindent
The acceleration $d\vec{v}_M / dt$ of a massive particle $M$ ("satellite") depends on the background mass-density $\rho$, the mass of the perturber $M$ and its velocity $v_M$. For more details on the function $F(v_M,\sigma)$, refer to Binney \& Tremaine (\cite{binney87}) Eq.\,(7-18). $C_f$ is a  scaling factor of order unity. As galaxies have density gradients, the force might point not exactly opposite to the velocity. Therefore we introduced an orthogonal component which is adjustable via an angle $\beta$. The Coulomb logarithm $\ln \Lambda$ is the relation between the maximum impact parameter $b_{\mathrm{max}}$ and the impact parameter $b_0$ that leads to a $90^\circ$ degree deflection. We found its best parameterisation to be mass and distance dependent -- cf.\,Eq.\,(\ref{petsch_eq_6}), where $M_\mathrm{halo}(r_M)$ is the mass of the host halo enclosed within the satellite's distance $r_M$.\\

\noindent
As our simulation uses static dark matter (DM) halos, the structural change due to the interaction (e.g. mass-loss and memory effect in the perturbed region) is not included.
As a first test we decided to directly implement mass-loss to our restricted {\sl N-}body code. At the time of the pericenter we are instantaneously reducing the mass of both interacting galaxies by the same factor. Certainly this is a very rude and heuristic approach, but it will already give a hint, if the shortcomings of the dynamical friction equation for equal mass merger can be reduced by including mass-loss.\\

Another important fact is, that the internal structure of the involved systems has a large impact on the merging process. E.g.\ the merging time is strongly correlated to the velocity dispersion. Therefore we exchanged our static isothermal halo by a static NFW (Navarro \etal \cite{navarro97}) halo.

\section{Results}

\begin{figure}[hb]\centering
\vspace{-0.2cm}
\rotatebox{-90}{\resizebox{!}{8cm}{%
\includegraphics[width=8cm]{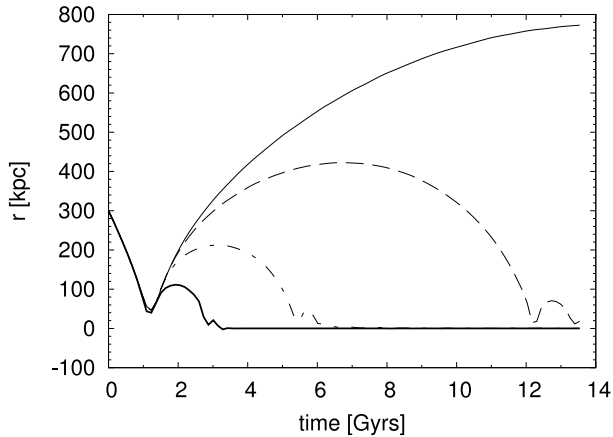}}}
\caption{Orbital decay of self-consistent equal mass mergers with different halo properties. Isothermal halos with different velocity dispersions (grey lines, decreasing order: solid, dashed, dash-dotted) are compared with a disk merger (Kuijken \& Dubinski \cite{kuijken95}), set up as one pro- and one retrograde disk (black solid line). All models have been set on an initially parabolic orbit with $r_\mathrm{halo}/r_\mathrm{min}=5$ and evolved with \textit{gyrfalcON} (Dehnen \cite{dehnen00}) tree-code.}
\label{petsch_fig_4a}
\end{figure}

\begin{figure}[ht]\centering
\vspace{-0.2cm}
\rotatebox{-90}{\resizebox{!}{8cm}{%
\includegraphics[width=8cm]{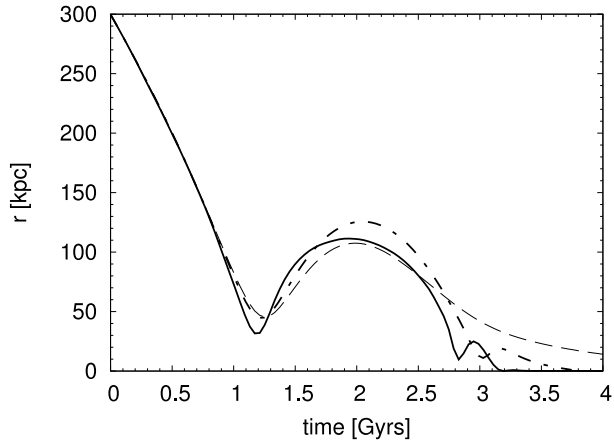}}}
\caption{Remodeling of the  radial decay of an equal mass disk merger (same as shown in Fig.\,\ref{petsch_fig_4a}, black solid line). Restricted models used a mass and distance dependent Coulomb logarithm. Isothermal halo, no mass-loss (grey dashed line); NFW halo, $1/3$ of the total mass is lost at each pericenter (black dash-dotted line).}
\label{petsch_fig_4b}
\end{figure}

It is shown in Fig.\,\ref{petsch_fig_4a}, that the internal structure of a halo has a strong impact on the merging time. If the velocity dispersion within an isothermal halo is reduced by a factor of $1.22$, the merging time can drop by a factor of up to $5$. If disk-like galaxies are used instead of pure isothermal halos, the merging time gets even smaller. In order to achieve reasonalbe merging times, we have decided to use NFW halos for future restricted N-body models, in addition they are able to describe the halo properties of a disk galaxy better than isothermal spheres. In order to quantify both processes from a stellar-dynamical point of view, we performed simulations with NFW profiles and the direct ad-hoc implementation of mass-loss described above. The results are shown in Fig.\,\ref{petsch_fig_4b}, for an appropriate strength of the frictional force, the radial decay can be modeled convincingly, though the implemented mass-loss model is very simple. At the moment we are improving the approach by measuring the mass-loss in self-consistent simulations and gauging the simplified approach.

%

\section{Conclusions and outlook}
In order to model observed major merger systems, we have implemented dynamical friction and a heuristic approach for the mass-loss. We have found, that NFW halos are more suitable to serve as host halos for our restricted {\sl N-}body models than isothermal halos (merging time, description of disk galaxy halo). Our results show, that implementation of mass-loss is necessary when modeling $1:1$ mergers in restricted {\sl N-}body models. At the moment we improve our implementation of mass-loss, which will be calibrated by measuring the mass-loss of self-consistent simulations.

\acknowledgements
This work was supported by the German Science Foundation (DFG) under the grant TH 511/9-1, which is part of the DFG priority program 1177.

\end{document}